\documentclass[aps,prd,showpacs]{revtex4}

\usepackage{latexsym}
\usepackage{amsfonts,amssymb}
\usepackage{graphicx,epsfig}
\usepackage{psfrag}

\newcommand{\be}{\begin{equation}}
\newcommand{\ee}{\end{equation}}
\def\bq{\begin{eqnarray}}
\def\eq{\end{eqnarray}}

\def\s{\sigma}

\def\Om{\Omega_{\rm m}}
\def\la{\lambda}
\def\tla{\Lambda}

\begin{document}

\title{Can brane cosmology with a vanishing $\Lambda$ explain the observations?}

\author{ R. G. Vishwakarma\footnote{e-mail: vishwa@iucaa.ernet.in} and
Parampreet Singh\footnote{e-mail: param@iucaa.ernet.in}}
\address{Inter-University Centre for Astronomy and Astrophysics,\\
Post Bag 4, Ganeshkhind, Pune-411 007, INDIA.}

\begin{abstract}
A plethora of models of the universe have been proposed in recent years claiming that
the present universe is accelerating, being driven by some hypothetical source with
negative pressure collectively known as \emph{dark energy} which though
do not appear to resemble any known form of matter tested in the laboratory.
These models are motivated by the high redshift supernovae Ia observations.
Though low density models, without dark energy, also appear to fit the SN Ia
data reasonably well, however, they are ruled out by the CMB observations.

In this paper, we present a warped brane model with an additional surface term of
brane curvature scalar in the action. This results in shifting the \emph{dynamical
curvature} of the model from its \emph{geometrical} counterpart,
which creates profound consequences.
Even for $\Lambda=0$, the low energy decelerating model successfully
explains the observed locations of the peaks in the angular power spectrum of CMB.
This model also fits the high redshift supernovae Ia observations,
taken together with the recently observed SN 1997ff at $z\approx 1.7$, 
very well. Additionally,
it also fits the data on the angular size and redshift of the compact radio sources
very well.
\end{abstract}

\pacs{9880, 9880E, 9880H, 0420}

\maketitle

\section{Introduction}
\noindent
In the past few years, there has been a spurt of activity in discovering models of the universe
in which the expansion is accelerating, fuelled by some self-interacting smooth
unclustered fluid with high negative pressure collectively known as
{\it dark energy} (for a recent review, see \cite{peebles}). These models
are mainly motivated by the high redshift SuperNovae (SN) Ia observations
which cannot be explained
in the framework of the canonical Einstein deSitter model (a model of the decelerating expansion
of the universe). The {\it dark energy}-models are also supported by
the recent measurements of the fluctuations in the power spectrum of the cosmic microwave
background (CMB) radiation
which appear to indicate that the universe is spatially flat and, hence,
$\Omega_{\rm total}\approx 1$ \cite{bernardis, cmbobs}.
However, the contribution form the total mass density, including the dark matter,
is only about one third: $\Omega_{\rm m0}\approx 0.3$,
which comes from the studies of the evolution of cluster
abundances with redshift, measurements of the power spectrum of large-scale
structure, analyses of measured peculiar velocities as they relate to the
observed matter distribution, and observations of the outflow of material
from voids \cite{turner1} (the subscript 0 denotes the value of the quantity at the present epoch).
Thus, the remaining two thirds of $\Omega_{\rm total}$ can easily be accounted by the
{\it dark energy}.

The simplest model for {\it dark energy} is the cosmological constant $\Lambda$,
which is though plagued with the so called the cosmological constant problem: why don't we see
the large vacuum energy density $\rho_{\rm v}\equiv \Lambda/8\pi G\approx 10^{76}$ GeV$^4$,
expected from particle physics which  is $\approx 10^{123}$ times larger than the value
predicted by the Friedmann equation? A phenomenological solution to this problem is
supplied by a dynamically decaying $\Lambda$ \cite{vishwa1, vishwa2, vishwa3}
or by an evolving large-scale scalar field, commonly known as {\it quintessence}, which can
produce negative pressure for a potential energy-dominated field \cite{peebles}.
A plethora of models of \emph{dark energy} has erupted from these ideas in
the past few years claiming that the present universe is accelerating \cite{peebles}.
It should be noted however that despite its consistency with
the observations, the nature of \emph{dark energy} is a mystery at present.
It does not seem to resemble any known form of matter tested in the laboratory.
As yet, we have no direct indication that it really exists.
In fact, a more natural value of the cosmological constant is zero (which could either be
due to some symmetry or due to a dynamical adjustment mechanism) rather than an
extremely small value but still non-zero.
There have been numerous suggestions that the apparent complications can be
eliminated by modifying the laws of gravity \cite{alternate}.

In this paper, we show that the present observations $-$ measurements of the angular
fluctuations in the power spectrum of the CMB, magnitude-redshift observations of the
high redshift SN Ia including SN 1997ff of $z\approx 1.7$, angular size-redshift observations
of the radio sources of size milliarcsecond $-$ can successfully be explained \emph{without
a $\Lambda$-term} and the universe is still decelerating. The background model, we consider
for this purpose, comes from the warped brane cosmology with an addition of a brane curvature
scalar in the action.
Higher dimensional (brane) models, inspired by the superstring theory solutions, are
acquiring attention. One possibility of great importance arising from these models is
that the fundamental Planck scale in the higher dimensions can be considerably smaller
than the usual Planck scale in our 4-dimensional spacetime. This would have profound
consequences for models of the very early universe.
For ready reference and also for completeness, we describe the model in the next section.
Observational and cosmological consequences of the model are studied in the following
sections.

\section{The model}
\noindent
In brane cosmology, the homogeneous, isotropic Robertson Walker (RW) universe can be envisioned
as a hyper surface embedded in the Schwarzschild anti-deSitter (AdS) bulk spacetime.
For the RW metric

\be
d s^2 =  - d\,  t^2 +  S^2(t)\,  \left(\frac{d r^2}{1 - k r^2} + r^2d \, \theta^2 + r^2\sin^2 \,
\theta \, d \, \phi^2 \right),\label{eq:metric}
\ee
the Israel junction conditions
\cite{israel} yield a Friedmann-like equation which has some additional terms
\cite{muk,kraus,bin}:
\be
H^2 = \frac{8 \pi G}{3} \, \rho \, \left(1 + \frac{\rho}{2 \s} \right) +
\frac{\la}{3} + \frac{C}{S^4} - \frac{k}{S^2} \label{eq:fosb},
\ee
where $S$ is the scale factor,
$G$ and $\la$ are respectively 4-D gravitational and
cosmological constants,  $\s$ is the brane tension and
$C$ is the mass parameter of the bulk black hole. The energy density
$\rho=\rho_{\rm m}+\rho_{\rm r}$ ($\rho_{\rm m}\equiv \rho_{\rm matter}$,
$\rho_{\rm r}\equiv \rho_{\rm radiation}$), where $\rho_{\rm m}$ has contributions from
baryonic as well as dark matter (discrepancies between the apparent baryon content of
galaxies and galaxy clusters and their dynamically inferred masses).
Equation (\ref{eq:fosb}) differs from the usual Friedmann equation of the standard cosmology in
the following two
terms. Firstly, the $\rho^2$ term which arises from the corrections in the
stress tensor, by imposing the junction conditions; and secondly, the
\emph{dark radiation} term $C/S^4$, which arises from the projection
of Weyl curvature of the bulk black hole on the brane and behaves like an
additional collisionless and isotropic massless component.
However, as we shall show later, these modifications die out at low redshifts.

If one considers quantum corrections to the bulk-brane action, then further
modifications to the Friedmann equation can be obtained.
One such correction, which is also needed for stress tensor regularization \cite{kraus1},
has been widely discussed in the brane world scenarios \cite{myung, collinsYuri, svd}.
It is achieved by adding the Ricci curvature scalar of the brane as a surface term in
the action. In warped brane models, this shifts the curvature parameter
$k$ ($=0,\pm 1$) of the RW metric to $k_{\rm eff}\equiv k-\alpha$ and equation
(\ref{eq:fosb}) leads to \cite{myung, svd}

\be
H^2 =   \frac{8 \pi G}{3} \, \rho \, \left(1 + \frac{\rho}{2 \s} \right) + \frac{\tla}{3}
+ \frac{\cal C}{S^4}  - \frac{(k - \alpha)}{S^2}\label{eq:nfeq}
\ee
where
\bq
\alpha &=& b \, \bigg[ \sqrt{\frac{\pi \, G \, \s}{3}} \, l - \frac{1}{16 \, \pi \, G \, \s \, l^2}\bigg] ,\\
\tla &=& \la - \frac{3 \, b}{l} \, \sqrt{3  \pi  G  \s} \, \bigg[\frac{8 \, \pi \, G \, \s \, l^2}{9} + \frac{1}{4  \pi  G  \s  l^2} - 1 \bigg] ,\\
{\cal C} &=& C \, \bigg[1 + b \left(\frac{3 - \beta^2}{3 \, \beta} \right)\bigg], ~~~~ \beta = 4 \, l \, \sqrt{\frac{\pi \, G \, \s}{3}}.
\eq
Here $b$ is a small dimensionless parameter through which the brane
curvature couples to the action and $l$ is the radius of curvature of the bulk spacetime.
Equation (\ref{eq:nfeq}) provides the model
which we shall be considering in this paper.
Some cosmological implications of this model have been studied earlier by considering
a non-zero $\Lambda$ \cite{svd}.

By interpreting the different energy density components in units of the critical density,
in the usual forms:
\be
 \Om \equiv \frac{8 \pi G}{3 H^2} \rho_{\rm m}, \,\hspace{0.25cm}
\Omega_{\rm r} \equiv \frac{8 \pi G}{3 H^2} \rho_{\rm r}, \, ~~
 \Omega_{\Lambda} \equiv \frac{\tla}{3 H^2}, \, ~~
\Omega_{\rm dr}\equiv \frac{{\cal C}}{S^4 H^2}, ~~
\Omega_{k_{\rm eff}} \equiv \frac{(k - \alpha)}{S^2 H^2}, \label{eq:omegas}
\ee

equation (\ref{eq:nfeq}) reduces to,
\bq
H^2(z) &=& H^2_0 \, \bigg[ \Omega_{{\rm m} 0} \, (1 + z)^3
\left\{1+ \frac{\rho_{\rm m0}}{2\s}(1+z)^3 + \frac{\rho_{{\rm r}0}}{\s}(1+z)^4\right\}\nonumber \\
&+&\Omega_{{\rm r} 0}(1+z)^4 \left\{1+ \frac{\rho_{\rm r0}}{2\s}(1+z)^4\right\}
+ \Omega_{\Lambda 0}+\Omega_{\rm dr 0}(1 + z)^4
-\Omega_{k_{\rm eff}0} \, (1 + z)^2 \bigg] \label{eq:ofe3}.
\eq
Constraints have been put on the brane tension $\s$ by requiring that
the relative corrections to the Newtonian gravity at short distances should be small.
This constrains $\s$  by $\s > 10^9$ GeV$^4$ \cite{maartens} giving
$(\rho_{{\rm r}0}/\s) < 10^{-61}$ and  $(\rho_{{\rm m}0}/\s) < 10^{-56}$. Thus the
terms containing $\s$ in equation (\ref{eq:ofe3}) are significant only for the
redshifts $z>10^{15}$ and equation (\ref{eq:ofe3}) reduces to the following
in the later epochs:

\be
H^2(z) = H^2_0 \, \bigg[ \Omega_{{\rm m} 0} \, (1 + z)^3
+ (\Omega_{{\rm r} 0}+\Omega_{\rm dr 0})(1 + z)^4
+ \Omega_{\Lambda 0}
-\Omega_{k_{\rm eff}0} \, (1 + z)^2 \bigg]. \label{eq:ofe4}
\ee
The \emph{dark radiation} term is constrained by
the big bang nucleosynthesis (BBN) by requiring that the change
in the expansion rate due to this term be sufficiently
small, so that an acceptable helium-4 abundance is produced.
This yields $ 0\leq \Omega_{\rm dr 0} \leq 0.11 \times \Omega_{r 0}$ \cite{ichiki}.
Note that the horizon condition of the black hole in the bulk requires that
$\Omega_{\rm dr}$ should be non-negative for the cases $k = 0$ and $1$
\cite{muk,kraus,s-ads}.
By using the energy momentum conservation, the deceleration parameter
$q\equiv -\ddot{S}/S H^2$ can be obtained from equation (\ref{eq:nfeq}) in the form

\be
q(t) = \frac{4\pi G}{3H^2} (\rho+3p)-\frac{\tla}{3H^2}+\frac{\cal C}{S^4H^2}
=\frac{\Omega_{\rm m}}{2}+\Omega_{\rm r}+\Omega_{\rm dr}-\Omega_\Lambda.\label{eq:decel}
\ee

The effective curvature index ($k-\alpha$) in equations (\ref{eq:nfeq}) and
(\ref{eq:ofe4}) now serves the role of the {\it dynamical curvature}, since the 
expansion dynamics
of the model, by using (\ref{eq:ofe4}) at $z=0$, can be written solely in terms of the source terms:

\be
H^2=H_0^2[\{1+z\Omega_{{\rm m}0}+z(2+z)(\Omega_{{\rm r}0}+\Omega_{{\rm dr}0})\}
(1+z)^2-z(2+z)\Omega_{\Lambda0}];\label{eq:fried}
\ee
and it does not have any term of $\alpha$.
The same is true for other quantities, like the deceleration parameter and the
expansion age of the universe, which do not have any term of $\alpha$.
However, the different distance measures do depend on $\alpha$ (at least
for the non-zero $k$, as will become clear in the following), through the {\it dynamical
curvature}. It should be noted that, unlike the standard cosmology, the {\it dynamical
curvature} of the model is different from its {\it geometrical curvature} $k$, which
has profound consequences in the model. To illustrate this point, we derive
different distance measures in the model.

If a light source of redshift $z$ is located at a radial coordinate distance
$r_1$, its {\it luminosity distance} $d_{\rm L}$, {\it angular diameter distance}
$d_{\rm A}$ and {\it proper motion distance} $d_{\rm M}$ are given by
\be
d_{\rm L} = (1 + z) S_0 ~r_1, ~ ~ d_{\rm A}=\frac{S_0 ~r_1}{(1+z)}, ~ ~ d_{\rm M}=S_0 ~r_1,
\label{eq:dist}
\ee
where $r_1$ can be calculated from the metric (\ref{eq:metric}), giving

\be
\frac{1}{S_0} \, \int_0^z \, \frac{d z'}{H(z')}=
\left\{ \begin{array}{ccl}
\vspace{0.2cm}
\sin^{-1}r_1,& \mbox{when}& k = 1 \\
\vspace{0.2cm}
r_1, &\mbox{when}& k = 0 \\
\sinh^{-1}r_1,& \mbox{when}& k = -1.
\end{array}\right. \label{eq:rdist}
\ee
It is worth noting that in deriving equations (\ref{eq:dist}, \ref{eq:rdist}), the only
assumption one needs
to make, is that of the validity of the RW metric given by equation (\ref{eq:metric})
and hence they hold for all the RW models.
The present {\it radius} of the universe $S_0$, appearing in
equations (\ref{eq:dist}) and (\ref{eq:rdist}) which measures the curvature of spacetime,
can be calculated from equation (\ref{eq:omegas}) as

\be
S_0 =\sqrt{\frac{(k-\alpha)}{(\Omega_{\rm total}-1)}} H_0^{-1},~ ~ ~
\Omega_{\rm total} \equiv
\Omega_{{\rm m}0}+\Omega_{{\rm r}0}+\Omega_{{\rm dr}0}+\Omega_{\Lambda 0}.\label{eq:szero}
\ee

An interesting case arises when the model is \emph{geometrically flat} ($k=0$).
More explicitly, this case can be represented by $\Omega_{{\rm m}0}+\Omega_{{\rm r}0}+
\Omega_{{\rm dr}0}+\Omega_{\Lambda 0}+\Omega_{\alpha 0}=1$, where the virtual $\Omega_{\alpha}$
is defined by $\Omega_{\alpha}\equiv \alpha/S^2H^2$. This is analogous to the case
$\Omega_{{\rm m}0}+\Omega_{{\rm r}0}+\Omega_{\Lambda 0}=1$ of the standard flat cosmology.
Since the expansion dynamics as well as the different distance measures 
do not depend on 
$\alpha$ in the $k=0$ model,
$\Omega_{{\rm m}0}$, $\Omega_{\Lambda 0}$, etc. can vary freely (by choosing suitable
$\Omega_{\alpha 0}$) and can produce an $\Omega_{\rm total}$ significantly different from 1
(unlike the standard cosmology: $\Omega_{\rm total}=1$).
This is, in fact, the only apparent difference (apart from $\Omega_{\rm dr}$) between the
present ($k=0$) model and the standard flat cosmology. In the latter case, the source
parameters are constrained to remain on the plane $\Omega_{{\rm m}0}+\Omega_{{\rm r}0}+\Omega_{\Lambda 0}=1$;
whereas in the former case, they are constrained by
$\Omega_{{\rm m}0}+\Omega_{{\rm r}0}+\Omega_{{\rm dr}0}+\Omega_{\Lambda 0}+\Omega_{\alpha 0}=1$.
The only other constraint the parameters have to satisfy is
$1+\Omega_{\rm m0}z+ z(2+z)\{\Omega_{\rm r0}+\Omega_{\rm dr0}-\Omega_{\Lambda0}/(1+z)^2\}\geq 0$,
coming from equation (\ref{eq:fried}). This gives more leverage to the parameters
$\Omega_{\rm m0}$ and $\Omega_{\Lambda_0}$ and hence a bigger parameter space is allowed
than in the flat standard cosmology.
Additionally for a given $\alpha$, one can still estimate $S_0$ (for the cases
$\Omega_{\rm total}\neq 1$)
in a \emph{geometrically flat} model through
$S_0 =\sqrt{\alpha/(1-\Omega_{\rm total})} H_0^{-1}$, which is otherwise not possible
in the standard cosmology. This implies that $\alpha$ is, respectively, positive or negative
according as $\Omega_{\rm total}<1$ or $\Omega_{\rm total}>1$.
For $\Omega_{\rm total}= 1$, the model reduces to the standard flat model.
In this paper, we shall keep ourselves limited to the case $k=0$. A more general
analysis will be done elsewhere \cite{later}.

We would like to mention about an alternative interpretations of the term $\alpha/S^2$
appearing in equation (\ref{eq:nfeq}). In fact, the $1/S^2$ fall off is a typical
characteristic of a `pseudo' source with equation of state
$\rho + \Sigma_i \, p_i = 0$ which corresponds to topological
defects, like cosmic strings and textures \cite{cs}. However, it neither contributes
to the deceleration parameter nor to the Hubble parameter, as we have seen earlier.
It would, therefore, be more appropriate to consider this term as a shift in the
\emph{dynamical curvature} of the standard cosmology and not as a source term.

\section{Observations of the fluctuations in the power spectrum of CMB }

\noindent
The small fluctuations (anisotropies) in the temperature of CMB offer a
glimpse of the epoch in the early universe when photons
decoupled from the cosmic plasma at $z_{\rm dec}=1100$.
Before this epoch, matter and radiation were tightly coupled and behaved like a single
fluid. At $z = 1100$, the temperature dropped sufficiently to let the protons
capture electrons to form neutral hydrogen and other light elements
(\emph{recombination}). As the electrons, which had trapped photons, disappeared
reducing the opacity for Thomson scattering, the
photons \emph{decoupled} (\emph{last scattered}) from matter.
The primordial quantum
fluctuations, in matter as well as radiation, themselves may have their origin
in the period of inflation. In the case of matter these fluctuations grew due to
gravitational instability forming the present structure. However, the radiation has
been streaming freely since decoupling. It has been highly redshifted, down to the micro
region, and carry the picture of the last
scattering surface left imprinted in the form of angular anisotropy in its intensity.

The initial fluctuations in the tightly coupled baryon-photon plasma oscillate at the
speed of sound driven by gravity, inertia of baryons and pressure from photons. This
continues until the recombination epoch. Physically these oscillations represent the
hot and cold spots on the fluid generated by compression and rarefaction by a standing
sound or acoustic wave.
Thus the wave which has a density maximum at the time of last scattering, corresponds
to a peak in the power spectrum. In the Legendre multipole space $\ell$, this corresponds
to the angle subtended by the sound horizon at the last scattering. Higher
harmonics of the principal oscillations, which have oscillated more than once,
correspond to secondary peaks. These locations of the peaks are very sensitive to
the variations in the parameters of the model and hence serve as a sensitive probe
to constrain the cosmological parameters and discriminate among various models
\cite{hu, doran}.

The angular power spectra of the temperature fluctuations in CMB have recently
been measured in many experiments, like, BOOMERANG, MAXIMA, DASI, CBI, etc.
\cite{bernardis, cmbobs}. Bernardis et al have recently measured the ranges of the
first three peaks from their improved analysis of a bigger sample of the
BOOMERANG data \cite{bernardis}:
$\ell_{\rm peak_1}:200-223$,
$\ell_{\rm peak_2}:509-561$,
$\ell_{\rm peak_3}:820-857$ at one sigma level; and
$\ell_{\rm peak_1}:183-223$,
$\ell_{\rm peak_2}:445-578$,
$\ell_{\rm peak_3}:750-879$ at two sigma level.
Though the uncertainties in these  $\ell_{\rm peak}$ values are large,
all the observations made so far agree at least on the location of the first peak
which has been measured comparatively more accurately.

The locations of the peaks are set by the acoustic scale $\ell_{\rm A}$, which
can be interpreted as the angle subtended by the sound horizon at the
last scattering surface. This angle (say, $\theta_{\rm A}$) is given by the ratio
of sound horizon to the distance (\emph{angular diameter distance}) of the last
scattering surface:

\be
\theta_{\rm A}=\frac{S(t_{\rm dec})\int_0^{t_{\rm dec}}c_{\rm s}\frac{dt}{S(t)}}
{d_{\rm A} (t_{\rm dec})},\label{eq:thetaa}
\ee
where  the speed of sound $c_{\rm s}$ in the plasma is given by $c_{\rm s}=1/\sqrt{3(1+R)}$
and $R\equiv3\rho_{\rm b}/4\rho_\gamma=3\Omega_{\rm b0}/[4\Omega_{\gamma0}(1+z)]$
corresponds to the ratio of baryon to photon density.
Acoustic scale $\ell_{\rm A}=\pi/\theta_{\rm A}$, in the $k=0$ model, then yields

\be
 \ell_{\rm A}=\pi
 \frac{\int_0^{z_{\rm dec}}\left[\{1+z\Omega_{{\rm m}0}+z(2+z)(
 \Omega_{{\rm r}0}+\Omega_{{\rm dr}0})\}(1+z)^2-z(2+z)\Omega_{\Lambda0}
 \right]^{-1/2}dz}
 {\int_{z_{\rm dec}}^\infty c_{\rm s}\left[\{1+z\Omega_{{\rm m}0}+z(2+z)(
 \Omega_{{\rm r}0}+\Omega_{{\rm dr}0})\}(1+z)^2-z(2+z)\Omega_{\Lambda0}
 \right]^{-1/2}dz}. \label{eq:la}
\ee
The location of $i$-th peak in the angular power spectrum is given by

\be
\ell_{\rm peak_i}=\ell_{\rm A}(i-\delta), \label{eq:lpeak}
\ee
where the phase shift $\delta$, caused by the plasma driving effect, is solely
determined by the pre-recombination physics and hence would not have any significant
contribution from the term containing $\alpha$ at that epoch. Thus one can 
safely
approximate $\delta$ (at least for the present quality of data)
with its value in the standard cosmology \cite{hu}, viz.

\be
\delta \approx 0.267 \left\{\frac{r(z_{\rm dec})}{0.3} \right\}^{0.1},
\ee
where $r(z_{\rm dec})\equiv \rho_{\rm r}(z_{\rm dec})/\rho_{\rm m}(z_{\rm dec})=
\Omega_{\rm r0}(1+z_{\rm dec})/\Omega_{\rm m0}$ is the ratio of radiation and matter
energy densities at the decoupling era.

Note that $\Omega_{\rm r0}$ gets contributions from photons (CMB) as well as neutrinos:
$\Omega_{\rm r}=\Omega_{\gamma}+\Omega_{\nu}$. The present photon contribution to the
radiation can be estimated by the CMB temperature $T_0=2.728$K. This gives
$\Omega_{\gamma0}\approx 2.48 ~ h^{-2}\times 10^{-5}$, where $h$ is the present value
of the Hubble parameter in units of 100 km s$^{-1}$ Mpc$^{-1}$.
The neutrino contribution follows from the assumption of 3 neutrino species, a standard
thermal history and and a negligible mass compared to its temperature \cite{hu2}.
This supplies $\Omega_{\nu0}\approx 1.7 ~ h^{-2}\times 10^{-5}$. We have already
mentioned that the allowed range of the \emph{dark radiation} coming from the BBN is
$[0-0.11]\times \Omega_{\rm r0}$. We consider an average value from this range, viz.,
$\Omega_{\rm dr0}=5.5\%$ of $\Omega_{\rm r0}$.
We find that the $\Lambda=0$ model fits the observed peaks very well for
a wide range of $\Omega_{\rm b0}$, $\Omega_{\rm m0}$ and $h$. Figure 1 shows the
ranges of $\Omega_{\rm m0}$ and $\Omega_{\rm b0}$ which produce the location of the
first peak $\ell_{\rm peak_1}$ in the observed region.
In order to check the robustness of the model, we vary $\Omega_{\rm dr0}$ at its
extreme limits, i.e., 0\% and 11\% of $\Omega_{\rm r0}$. The resulting models also fit
the observed peak locations very well. For example, for the values $\Omega_{\rm m0}=0.3$,
$\Omega_{\rm b0}=0.05$ and $h=0.65$, the different choices of $\Omega_{\rm dr0}$ yield
the following.
(i) $\Omega_{\rm dr0}=0$:
$\ell_{\rm peak_1}=197$,
$\ell_{\rm peak_2}=467$,
$\ell_{\rm peak_3}=737$;
(ii) $\Omega_{\rm dr0}=0.055\times\Omega_{\rm r0}$:
$\ell_{\rm peak_1}=199$,
$\ell_{\rm peak_2}=473$,
$\ell_{\rm peak_3}=748$;
(iii) $\Omega_{\rm dr0}=0.11\times\Omega_{\rm r0}$:
$\ell_{\rm peak_1}=201$,
$\ell_{\rm peak_2}=479$,
$\ell_{\rm peak_3}=756$.

It is obvious from the above that the influence of \emph{dark radiation} on the
locations of the peaks is to shift them towards higher $\ell$ values. This result is also
consistent with other investigations \cite{ichiki}. One also notices that some
of the above-mentioned $\ell_{\rm peak}$ values are towards the lower boundary of
the observations, especially when compared with 1-sigma observations. However,
they can be increased either by decreasing $h$ or by increasing $\Omega_{\rm b0}$.
For example, even with $\Omega_{\rm dr0}=0$, the values $\Omega_{\rm m0}=0.3$,
$\Omega_{\rm b0}=0.08$ and $h=0.6$ yield
$\ell_{\rm peak_1}=210$,
$\ell_{\rm peak_2}=502$,
$\ell_{\rm peak_3}=793$.

Just for comparison, it may be noted that the open model $\Lambda=0$, $\Omega_{\rm m0}=0.3$,
$\Omega_{\rm b0}=0.05$ with $h=0.65$
 in the standard cosmology ($\alpha=0=\Omega_{\rm dr0}$) yields
$\ell_{\rm peak_1}=425$,
$\ell_{\rm peak_2}=1010$,
$\ell_{\rm peak_3}=1594$, which is clearly ruled out by the observations.
The favoured standard model: $\Omega_{\rm m0}=1-\Omega_{\Lambda0}=0.35$,
for $\Omega_{\rm b0}=0.05$ and $h=0.65$, gives
$\ell_{\rm peak_1}=222$,
$\ell_{\rm peak_2}=524$,
$\ell_{\rm peak_3}=827$.

\begin{figure}[tbh!]
\epsfig{figure=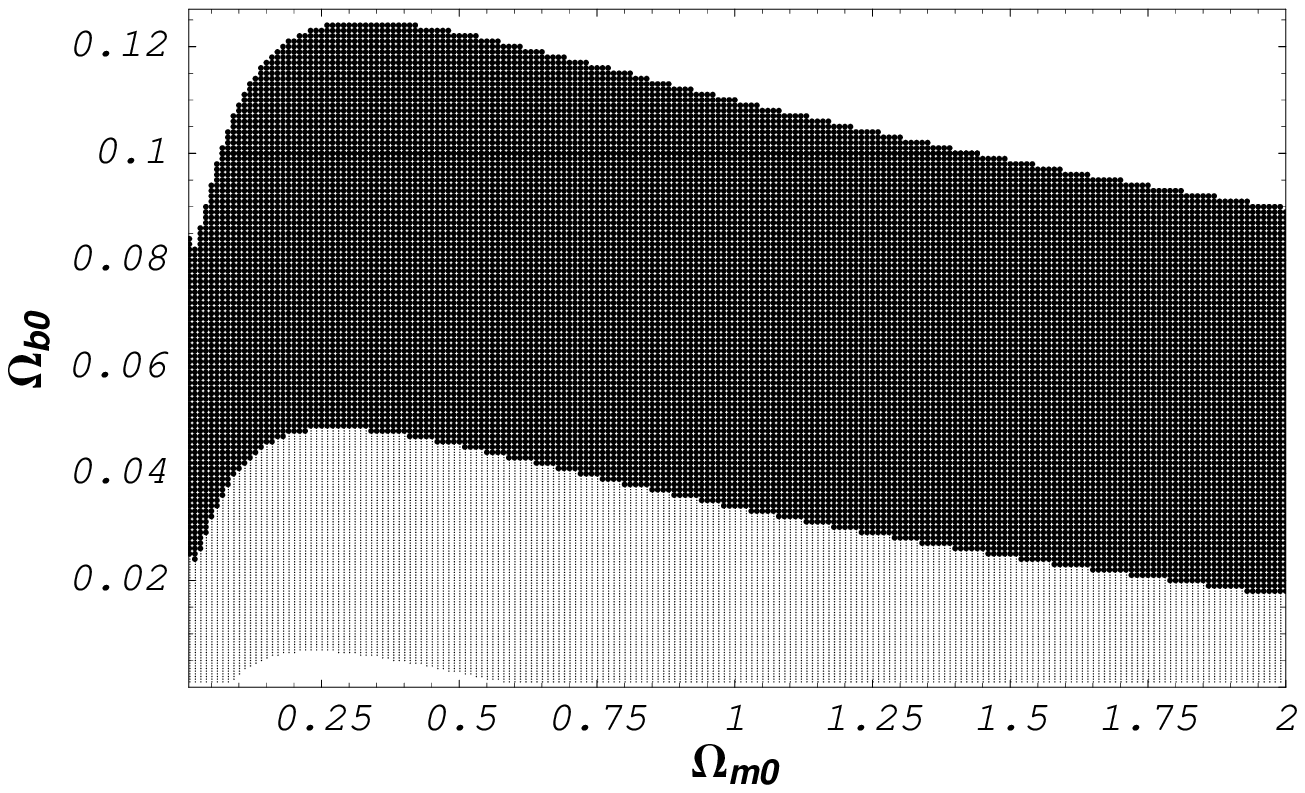,height=9.5in,width=7in,angle=0}
{\caption{\small The allowed regions by the CMB at 1 and 2 sigma levels
are shown in the $\Omega_{\rm m0}-\Omega_{\rm b0}$ plane in the $\Lambda=0$
cosmology, which produce the first peak
in the ranges, respectively, $200-223$ (dark-shaded region) and $183-223$ (light-shaded $+$
dark-shaded regions). We have considered $\Omega_{\rm dr0}=0.055 \times\Omega_{\rm r0}$ and
$h=0.65$.}}
\end{figure}

\section{Supernovae Ia observations}

\noindent
SuperNovae (SN) Ia, which are thought to be thermonuclear explosions of
carbon-oxygen white dwarfs \cite{trimble},
are almost universally regarded as {\it standard candles} and can be
observed at cosmological scales since they have high enough absolute
luminosity. As the {\it luminosity distance} $d_{\rm L}$ depends
sensitively on the spatial curvature and the expansion dynamics
of the models, the magnitude-redshift ($m$-$z$) relation for the
high redshift SN Ia can be used as a potential
test for cosmological models and provides a useful tool
to constrain the parameters of the models.

In order to fit to our model, we consider the data
on the redshift and magnitude of a sample of 54 SN Ia
considered by Perlmutter et al (excluding 6 outliers from the full
sample of 60 SN) \cite{perlmutter},
together with SN 1997ff at $z=1.755$, the highest redshift supernova observed
so far \cite{riess1}. It has been confirmed now that SN 1997ff has been
observed on the brighter side due to gravitational lensing from the galaxies
lying on the line of sight \cite{riess2}. After correction for lensing, the revised
magnitude of this SN is $m^{\rm eff}=26.02\pm 0.34$ \cite{riess2} which we shall use in our
calculations.

The standard definition of magnitude $m(z)$ for the RW spacetime (\ref{eq:metric})
is given by \cite{weinberg}

\be
m (z) = {\cal M} + 5 \,\log [{\cal D}_{\rm L}(z)], \label{eq:mageq}
\ee
where ${\cal M} \equiv (M - 5 \,\log H_0 +$ constant) is the \emph{Hubble constant-free}
absolute luminosity, $M$ is the absolute luminosity of the source
and  ${\cal D}_{\rm L} \equiv H_0 d_{\rm L}$ is the dimensionless luminosity
distance. The luminosity distance $d_{\rm L}$ in the model is supplied by equations
(\ref{eq:fried}$-$\ref{eq:rdist}). It has been shown in \cite{svd} that
the {\it dark radiation} term can be safely neglected while considering the present
SN Ia data. Moreover, the radiation terms can be neglected even at $z=3.8$ (which
corresponds to the highest redshift quasar we shall be considering in the next section)
since the contribution of the term $\Omega_{\rm r0}+\Omega_{\rm dr0}$ at $z=3.8$
(even when $\Omega_{\rm dr0}$ is at its highest limit) in equation (\ref{eq:ofe4})
is only about 0.2\% of the contribution from the term $\Omega_{\rm m0}$ with
$\Omega_{\rm m0}=0.3$.
The $\chi^2$ value is calculated from its usual definition

\be
\chi^2 = \sum_{i = 1}^{55} \,\left[ \frac{m_i^{\rm eff} - m(z_i)}
{\delta m_i^{{\rm eff}}}\right]^2,
\ee
where $m_i^{\rm eff}$ refers to the effective
magnitude of the $i$th SN which has been corrected by the lightcurve
width-luminosity correction, galactic extinction and the K-correction from
the differences of the R- and B-band filters and the dispersion
$\delta m_i^{\rm eff}$ is the uncertainty in $m_i^{\rm eff}$.

There seems to be a general impression that the high redshift SN observations rule out
the decelerating models. However, this is not true even in the standard cosmology, as
we shall see in the following. Equation (\ref{eq:decel}) suggests that the present
universe will be accelerating if $\Omega_{\Lambda0}>\Omega_{\rm m0}/2$.
Thus the best-fitting standard model to this sample, which is obtained for the values
$\Omega_{\rm m0}=0.7$, $\Omega_{\Lambda0}=1.2$, is of course accelerating. The
$\Delta\chi^2$ ($\chi^2$ per degrees of freedom), for this fitting, is obtained as
1.09 and the
probability $Q$, which measures the goodness of fit \cite{press}, is obtained as 29.7\%
which represent an excellent fit. Even the best-fitting flat model in the standard
cosmology is an accelerating one, which is obtained as
$\Omega_{\rm m0}=1-\Omega_{\Lambda0}=0.3$ with $\Delta\chi^2=1.11$ and $Q=26.6$\%.
Though the canonical Einstein-deSitter model ($\Omega_{\rm m0}=1-\Omega_{\Lambda0}=1$) has
a poor fit: $\Delta\chi^2=1.72$ and $Q=0.08$\% and is ruled out, however, the open
models with low $\Omega_{\rm m0}$ have reasonable fits. For example, the decelerating
(standard cosmological) models
$\Omega_{\rm m0}=0.2$ with $\Lambda=0$ ($\Delta\chi^2=1.24$, $Q=11.4$\%);
$\Omega_{\rm m0}=0.3$ with $\Lambda=0$ ($\Delta\chi^2=1.27$, $Q=8.5$\%);
$\Omega_{\rm m0}=0.4$ with $\Lambda=0$ ($\Delta\chi^2=1.32$, $Q=5.7$\%), etc.,
have reasonable fittings and by no means are rejectable. Open decelerating models
with non-zero $\Lambda$ ($0<\Omega_{\Lambda0}<\Omega_{\rm m0}/2$) have even better fit.

In the present (geometrically flat) brane model with $\Lambda=0$, $\chi^2$ decreases for
lower values of $\Omega_{\rm m0}$ (as in the standard open cosmology) giving the best-fitting
solution for $\Omega_{\rm m0}=-0.3$ (with $\Delta\chi^2=1.19$, $Q=16.5\%$), which is though
unphysical. Thus a `physically viable'
best-fitting solution can be regarded as $\Omega_{\rm m0}=0$, which gives
$\Delta\chi^2=1.22$ with $Q=12.4$\%, representing a reasonably good fit,
though not as good as the best fit in the standard cosmology.
The allowed regions by the data in the $\Omega_{\rm m0}-{\cal M}$
plane are plotted in Figure 2 at 1, 2 and 3 sigma levels.

\begin{figure}[tbh!]
\epsfig{figure=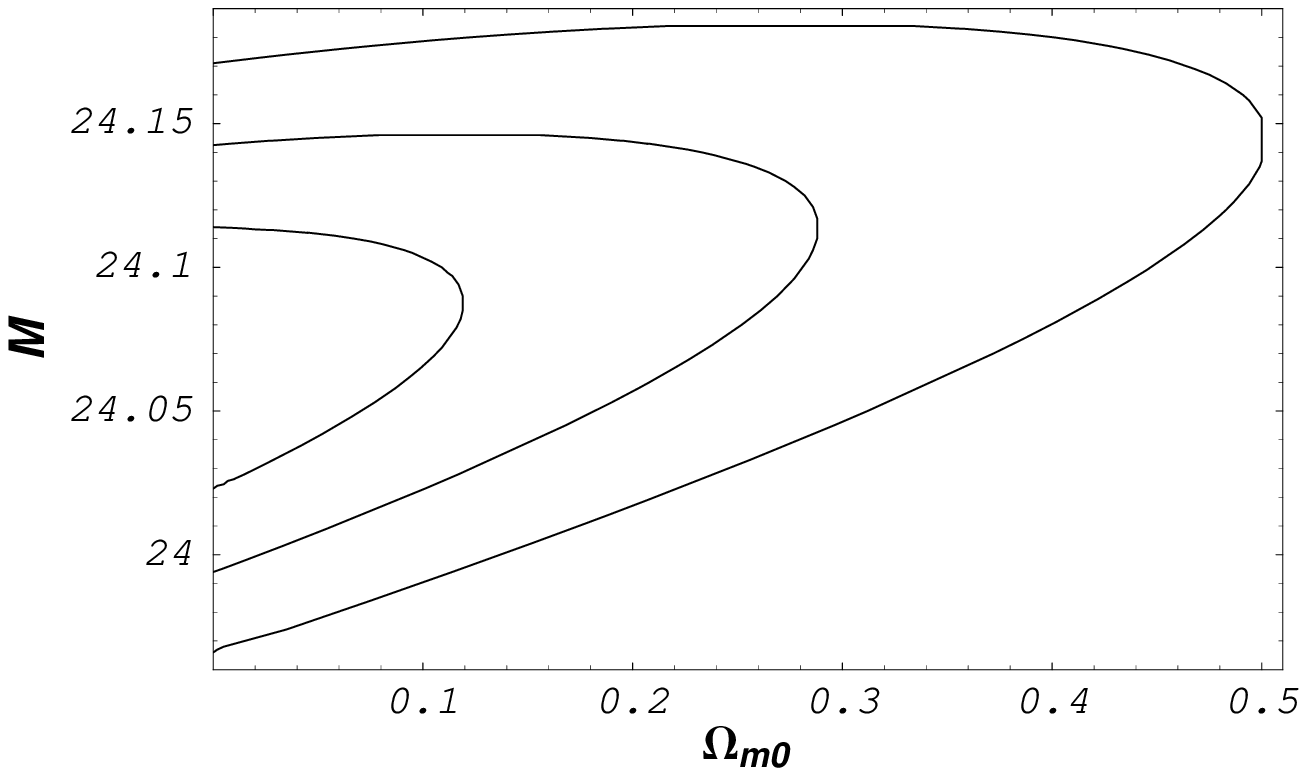,height=9.5in,width=7in,angle=0}
{\caption{\small The allowed regions by the SN Ia observations are shown in the
$\Omega_{\rm m0}-{\cal M}$ plane for $\Lambda=0$ cosmology 
($\cal M\equiv$ the \emph{Hubble constant-free} absolute luminosity).
The ellipses, in the order of increasing size, correspond to respectively 1, 2 and
 3 sigma levels.}}
\end{figure}

\section{Angular size-redshift data for high redshift compact radio sources}

\noindent
The radio sources are attractive for the
test of cosmological models because they are usually identified with
quasars which, in general, have high redshifts, so the models may be
more easily distinguished than with the extended double-lobed sources.
Kellermann \cite{kellerman} showed, by using a sample of 79 such sources, that the
resulting  angular size-redshift ($\Theta$-$z$) relation was cosmologically credible.
Jackson and Dodgson \cite{jackson1} used his data to test the standard cosmology with a
cosmological constant.
Later on, a more extensive exercise was carried out by Jackson and Dodgson
\cite{jackson2} by considering a bigger sample of 256 ultracompact sources
selected from the compilation from Gurvits \cite{gurvits1}. It is, in fact,
this data set we are considering to test the present model.
Although this data set has been extended recently by Gurvits et al
\cite{gurvits2}, however, the modified sample becomes highly inhomogeneous
as it compiles many different data sets obtained by many different
observers using different instrumentations and imaging techniques.
Hence its credibility is lost and it can fit any model \cite{vishwa2}.

The original data set of Jackson and Dodgson \cite{jackson2} (which we are considering
here) was selected from a bigger sample of 337 sources, out of which Jackson and
Dodgson selected the sources with $z$ in the range 0.5 to 3.8.
These objects, 256 in number, are ultra compact radio sources having angular
sizes of the order of a few milliarcseconds. They are deeply embedded in
the galactic nuclei and have very short life time compared with the
age of the universe. Thus they are expected to be free from the
evolutionary effects and hence may be treated as {\it standard rods},
at least in the statistical sense.
These sources are distributed into 16 redshift bins, each bin containing
16 sources. This compilation has recently been used by many authors to test
different cosmological models \cite{vishwa3, ban}.

In order to fit the data to the model, we derive the $\Theta$-$z$ relation in the
following. The (apparant) angular size $\Theta$ of a source, whose
proper diameter is $d$, is given by

\begin{equation}
\Theta (z)=\frac{0.0688 ~ d ~ h}{{\cal D}_A(z)} ~\mbox{milliarcseconds},
\end{equation}
where ${\cal D}_{\rm A} \equiv H_0 d_{\rm A}$ is the dimensionless angular
diameter distance and $d$ is measured in pc. In the present model, $d_{\rm A}$
can be obtained from equations (\ref{eq:fried}$-$\ref{eq:rdist}).
We fix $\Lambda=0$ and calculate the
theoretical $\Theta(z)$ for a wide range of parameters
$\Omega_{\rm m0}$ and $dh$. $\chi^2$ is computed from

\begin{equation}
\chi^2 =\sum_{i=1}^{16} \left[\frac{\Theta_i -
\Theta(z_i)}{\delta\Theta_i}\right]^2,
\end{equation}
which measures the agreement between the theoretical $\Theta(z_i)$ and the
observed value $\Theta_i$ with the  errors $\delta\Theta_i$
of the $i$th bin in the sample.

Here also $\chi^2$ decreases for lower values of $\Omega_{\rm m0}$ giving the best-fitting
solution for $\Omega_{\rm m0}=-0.15$ with $\Delta\chi^2=0.94$, $Q=51.9\%$, which is though
not interesting.
The `viable' best-fitting solution for $\Omega_{\rm m0}=0$ yields $\Delta\chi^2=0.95$
with $Q=50.4\%$, which represents an excellent fit.
Figures 3 shows the allowed regions by the data in the parameter space
$\Omega_{\rm m0}-dh$ at 1, 2 and 3 sigma levels.

\begin{figure}[tbh!]
\epsfig{figure=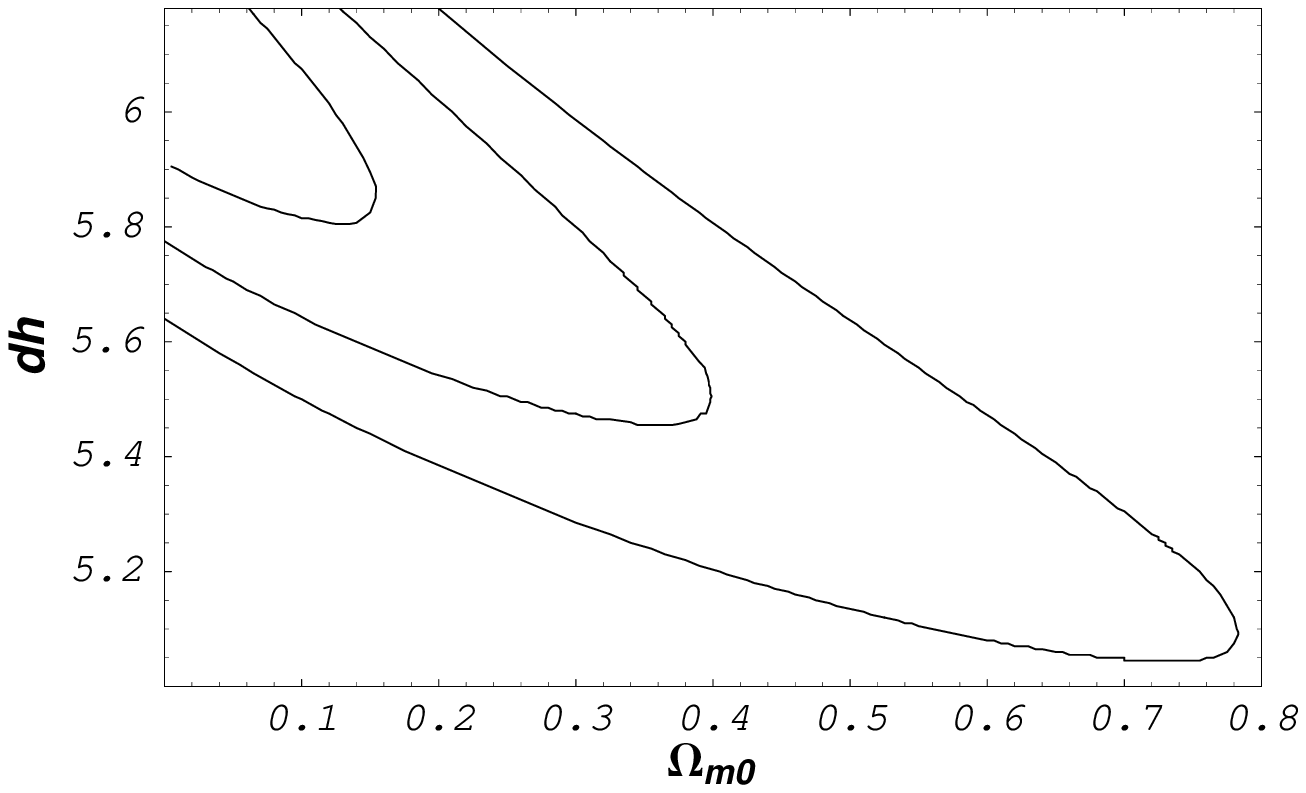,height=9.5in,width=7in,angle=0}
{\caption{\small The allowed regions by the compact radio sources data are shown in the
$\Omega_{\rm m0}-d h$ plane for $\Lambda=0$ cosmology (where $d$ is measured in pc).
 The ellipses, in the order of increasing size, correspond to respectively 1, 2 and
 3 sigma levels.}}
\end{figure}

\section{Conclusion}
\noindent
The best-fitting models to the high redshift supernovae Ia observations predict
an accelerating expansion of the universe driven by some unknown smooth fluid with negative
pressure, like the cosmological constant $\Lambda$. Although the open low energy
standard models without this strange fluid show reasonable fit to the SN Ia data, they
are ruled out by the recent CMB observations which predict that the universe is
spatially flat.
This has given rise to a plethora of models of the hypothetical
`\emph{dark energy}' (variants of $\Lambda$) which
do not seem to resemble any known form of matter tested in the laboratory.

In this paper, we have presented a warped brane model wherein the addition of the
brane curvature scalar as a surface term in the action results in shifting the \emph{geometrical
curvature} index $k$ of the standard cosmology by a constant $\alpha$. Though $\alpha$
does not enter directly into the expressions for the Hubble parameter and the different distances
measured in the geometrically flat ($k=0$) model, however
it helps $\Omega_{\rm total}$ to assume values different from 1,
which has profound consequences. Even for $\Lambda=0$, the low energy model successfully
explains the observed locations of the peaks in the angular power spectrum of CMB.
The model also fits the high redshift supernovae Ia observations,
taken together with the recently observed SN 1997ff at $z\approx 1.7$. Additionally,
it also fits the data on the angular size and redshift of the compact radio sources
very well.

Though $\alpha$ and $S_0$ do not enter into the dynamics of the (geometrically flat)
model, we can still extract some constraint on the sign of $\alpha$ from the observations,
which can be done through the relation $\alpha/S_0^2H_0^2=1-\Omega_{\rm total}$.
Obviously, the SN Ia and the radio sources data, which favour low density models
(see Figures 2 and 3) predict a positive $\alpha$. However, the CMB observations
are consistent with either sign of $\alpha$ (see Figure 1) for the available accuracy
of the data. This degeneracy might be removed by the future observations from the
MAP and PLANCK experiment which are expected to give more accurate measurements.

It is interesting to note that the low $\Omega_{\rm m0}$ models, which are consistent
with all the three observations we have considered, are also consistent with the dynamical
measurements of the energy density. Moreover, they are also consistent with the age of the
oldest objects detected so far, e.g., the globular clusters of age $t_{\rm GC}=12.5
\pm 1.2$ Gyr \cite{globul}. In this connection, we notice that the
recent measurements of $H_0$ favour a value
in the range (61 $-$ 65) km s$^{-1}$ Mpc$^{-1}$ \cite{hvalue1},
setting  the age of the universe $t_0$ in the following ranges. $\Omega_{\rm m0}=0.3$:
$t_0\in (12.2 - 13.0)$ Gyr; $\Omega_{\rm m0}=0.2$: $t_0 \in(12.7 - 13.6)$ Gyr,
which are consistent with the above-mentioned $t_{\rm GC}$.
A large number of other methods appear to converging on a value
of $H_0$ in the range (60 $-$ 80) km s$^{-1}$ Mpc$^{-1}$ \cite{hvalue2}.
An average value of $H_0=65$ km s$^{-1}$ Mpc$^{-1}$ from this range constrains
$\Omega_{\rm m0}$ of the model by $\Omega_{\rm m0}\leq 0.34$
to give $t_0 \geq 12$ Gyr.

\medskip
\noindent
\emph{Acknowledgements:} We thank Naresh Dadhich, Wayne Hu, Roy Maartens,
Jayant Narlikar, T Padmanabhan and Kandaswamy Subramanian for useful comments and discussions.
Thanks are also due to Paolo de Bernardis for sending the observed ranges
of the CMB peaks. RGV thanks DAE for
his Homi Bhabha postdoctoral fellowship and PS thanks CSIR for his
research fellowship.


\begin{references}




\bibitem{peebles} Peebles P. J. E. and Ratra Bharat, astr0-ph/0207347;\\
Sahni V. and Starobinsky A. 2000 Int. J. Mod. Phys. D {\bf 9} 373.


\bibitem{bernardis} de Bernardis P. et al 2002 Astrophys. J. {\bf 564}, 559.

\bibitem{cmbobs} de Bernardis P. et al 2000 Nature {\bf 404}, 955; \\
Lee A. T. et al 2001 Astrophys. J. {\bf 561}, L1; \\
Halverson N. W. et al 2002 Astrophys. J. {\bf 568}, 38;\\
Sievers J. L. et al, astro-ph/0205387.

\bibitem{turner1} Turner M. S.,  astro-ph/0106035.

\bibitem{vishwa1} Vishwakarma, R. G. 2002  Class. Quantum Grav. {\bf 19}, 4747;\\
 Vishwakarma, R. G. 2002 MNRAS, {\bf 331}, 776;\\
 Vishwakarma, R. G. 2001  Gen. Relativ. Grav. {\bf 33}, 1973.

\bibitem{vishwa2}  Vishwakarma, R. G. 2001  Class. Quantum Grav. {\bf 18},1159.

\bibitem{vishwa3} Vishwakarma, R. G. 2000  Class. Quantum Grav. {\bf 17}, 3833.


\bibitem{alternate}
Narlikar J. V. and Padmanabhan T. 2001 Ann. Rev. Astron. Astrophys. {\bf 39}, 211;\\
Boisseau B, Esposito-Farese G., Polarski D. and Starobinsky A. A.
2000 Phys. Rev. Lett. {\bf 85}, 2236;\\
Esposito-Farese G. and Polarski D. 2001 Phys. Rev. D {\bf 63}, 063504; \\
Gaztanaga E. and Lobo A. 2001 Ap. J. {\bf 548}, 47;\\
Narlikar J. V., Vishwakarma, R. G., Hajian Amir,
Souradeep Tarun, Burbidge G. and Hoyle F. 2003 Astrophys. J. {\bf 585}, 1
(astro-ph/0211036).\\
Bayin Selcuk S. 2002 IJMPD {\bf 11}, 1523 (astro-ph/0211097).


\bibitem{israel} Israel W. 1966 Nuov. Cim. B {\bf 44}, 1.

\bibitem{muk} Mukohyama S., Shiromizu T. and Maeda K. 2000 Phys. Rev. D {\bf 62}, 024028.

\bibitem{kraus} Kraus P. 1999 JHEP {\bf 9912}, 011.

\bibitem{bin} Binetruy P., Deffayet C., Ellwanger U. and Langlois D. 2000 Phys. Lett. B
{\bf 477}, 285.

\bibitem{kraus1} Balasubramanian V and Kraus P. 1999 Comm. Math. Phys. {\bf 208}, 413.

\bibitem{myung} Kim N. J., Lee H. W. and Myung Y. S. 2001 Phys. Lett. B {\bf 504}, 323.

\bibitem{collinsYuri} Collins H. and Holdom B. 2000 Phys. Rev. D {\bf 62}, 105009;\\
Shtanov Yu. V., hep-th/0005193.

\bibitem{svd} Singh P., Vishwakarma R. G. and Dadhich N., hep-th/0206193.

\bibitem{maartens} Maartens R. et al 2000 Phys.Rev. D {\bf 62}, 041301.

\bibitem{ichiki} Ichiki K. et al 2002 Phys. Rev. D {\bf 66}, 043521.

\bibitem{s-ads} Singh P. and Dadhich N., hep-th/0204190;\\
 Singh P. and Dadhich N., hep-th/0208080.

\bibitem{later} Vishwakarma R. G. and Singh P., \emph{In preparation}.

\bibitem{cs} Vilenkin A. and Shellard E. P. S. 1995 \emph{Cosmic strings
and other topological defects}, (Cambridge U. Press);\\
Dadhich N. and Narayan N. 1998 Gen. Relativ. Grav. {\bf 30}, 1133;\\
Dadhich N. and Patel L. K. 1999 Pramana {\bf 52}, 359;\\
Babul A. et al 1987 Astrophys. J. {\bf 316}, L49;\\
Gregory R. 1989 Phys. Rev. D. {\bf 29}, 2108;\\
Dabrowski M. P. 1989 Astron. J. {\bf 97}, 978.

\bibitem{hu} Hu W., Fukugita M., Zaldarriaga M. and Tegmark M. 2001 Astrophys. J.
{\bf 549}, 669.

\bibitem{doran} Doran M. and Lilley M. 2002 MNRAS {\bf 330}, 965;\\
Doran M. and Lilley M., Schwindt J. and Wetterich C. 2001 Astrophys. J. {\bf 559}, 501.

\bibitem{hu2} Hu Wayne and Dodelson Scott 2002 Ann. Rev. Astron. Astrophys.

\bibitem{trimble} Trimble V. 1983 Rev. Mod. Phys. {\bf 54}, 1183 \\
Woosley S. E.and Weaver T. A. 1986 ARA \& A {\bf 24}, 205.

\bibitem{perlmutter} Perlmutter S., et al 1999 Astrophys. J. {\bf 517}, 565.

\bibitem{riess1} Riess A. G., et al 2001 Astrophys. J. {\bf 560}, 49.

\bibitem{riess2} Narciso B., et al, astro-ph/0207097.

\bibitem{weinberg} Weinberg S. 1972 \emph{Gravitation and Cosmology},
(New York, John Wiley).

\bibitem{press} Press W., Teukolsky S., Vetterling W. and Flannery B. 1992
\emph{Numerical Recipes} chapter 15, (Cambridge University Press).

\bibitem{kellerman} Kellermann K I 1993 Nat. {\bf 361} 134.

\bibitem{jackson1} Jackson J C and Dodgson M 1996 MNRAS {\bf 278} 603.

\bibitem{jackson2} Jackson J C and Dodgson M 1997 MNRAS {\bf 285} 806.

\bibitem{gurvits1} Gurvits L I 1994 Astrophys. J. {\bf 425} 442.

\bibitem{gurvits2} Gurvits L I, Kellermann K I and Frey S. 1999 Astron. Astrophys.
{\bf 342} 378.

\bibitem{ban} Banerjee S K and Narlikar J V 1999 MNRAS {\bf 307} 73.

\bibitem{globul} Cayrel R., et al 2001 Nature, {\bf 409}, 691.\\
Gnedin O. Y., Lahav O. and Rees M. J., astro-ph/0108034.

\bibitem{hvalue1} Passnacht C. D. et al astro-ph/0208420.

\bibitem{hvalue2} Freedman W. B. 2000 Phys. Rept. {\bf 333}, 13.




\end{references}
\end{document}